\documentclass[]{article}
\newcommand{\bb}{\begin{eqnarray}}
\newcommand{\ee}{\end{eqnarray}}
\begin{document}
\title{An improved estimate of black hole entropy in the 
quantum geometry approach}
\author{A. Ghosh\thanks{amitg@theory.saha.ernet.in}
{} and P. Mitra\thanks{mitra@theory.saha.ernet.in}\\
Saha Institute of Nuclear Physics\\ 1/AF Bidhannagar\\
Calcutta 700064, India}
%\date{gr-qc/0411035}
\date{}
\maketitle
%\widetext
\begin{abstract}
A proper counting of states for black holes in the quantum geometry
approach shows that the dominant configuration for spins are distributions
that include spins exceeding one-half at the punctures. This
raises the value of the Immirzi parameter and the black hole
entropy. However, the coefficient of the logarithmic correction remains
-1/2 as before.
\end{abstract}
%\bigskip\bigskip\bigskip\bigskip
%\pacs{PACS Nos. 11.30.Er, 12.38.Aw}
%\narrowtext

The quantum geometry approach to a quantum theory of gravity is reasonably
well established now: see \cite{nicolai} for reviews.
In \cite{ash} a general framework for the calculation of black hole entropy 
in this approach was proposed. A lower bound 
for the entropy was worked out on the basis of the association
of spin one-half to each {\it puncture}
and found to be proportional to the area of the
horizon. The proportionality constant involves what is known as the
Immirzi parameter, which can be chosen so that the entropy becomes
a quarter of the area.

Recently, this lower bound was sharpened in \cite{lowerbound} to
include a logarithmic correction $-\frac{1}{2}\ln A$. Subsequently,
it was found \cite{km} that the dominant term in the entropy
is somewhat higher by taking spins higher than one-half
into account, though the logarithmic
correction is unaffected in this calculation. In the present note we
investigate the modification of the lower bound of \cite{lowerbound} in
view of this development and are led to a 
further increase in the leading term.

Let a generic configuration have $s_j$ punctures with spin $j, j=
1/2,1,3/2,2,...$. Note
that \bb 2\sum_js_j\sqrt{j(j+1)}=A\;,\label{areac}\ee where $A$ is
the horizon area in units where $4\pi\gamma\ell_P^2=1$, $\gamma$
being the Immirzi parameter and $\ell_P$ the Planck length. 
Following \cite{ash}, we shall treat the
punctures as distinguishable, see also \cite{rov}. We shall count states of 
the physical
Hilbert space considering both $j$ and its projection $m$ as quantum numbers. 
The difference between this procedure and the calculation carried out  
in \cite{km} will be commented on later.
%These authors have defined a quantum isolated horizon to have states 
%labelled {\em only} by $m$ quantum numbers, as opposed to both $j$ and $m$ 
%advocated here but in fact also considered and
%discussed in \cite{ash} to argue in favour of the {\it robustness} of the 
%counting. Here, we consider that possibility more seriously and define quantum
%isolated horizon states to be labelled by both the relevant quantum numbers. 

If we ignore the {\it zero spin projection constraint}
($\sum m=0$, the sum extending over all punctures)
initially, the total number of states
is given by \bb N={(\sum_j s_j)!\over \prod_j
s_j!}\prod_j (2j+1)^{s_j},\ee 
where one has to sum over all nonnegative
$s_j$ consistent with the given value of $A$. We will estimate the
sum by maximizing the above expression with respect to the variables
$s_j$ subject to a fixed value of $A$.

Using Stirling's formula, we see that \bb\ln N=\sum_j s_j\Big[\ln
(2j+1)-\ln s_j\Big] + (\sum_j s_j)\ln (\sum_j s_j)\;.\ee Hence,
\bb\delta\ln N=\sum_j\delta s_j\Big[\ln (2j+1)-\ln
s_j+\ln\sum_ks_k\Big]\;,\ee so that with some Lagrange multiplier
$\lambda$ to implement the area constraint, we can set \bb\ln
(2j+1)-\ln s_j+\ln\sum_ks_k-\lambda\sqrt{j(j+1)}=0\;.\ee Thus, \bb
s_j=(2j+1)\exp\Big[-\lambda\sqrt{j(j+1)}\Big]\sum_ks_k\;.\ee Summing
over $j$, we obtain the relation \bb\sum_j(2j+1)\exp\Big
[-\lambda\sqrt{j(j+1)}\Big]=1\;,\label{lambda}\ee which determines
$\lambda\simeq 1.72$. 
It may be mentioned that (\ref{lambda}) was noted as a
mathematical possibility in \cite{km}, and was derived with a somewhat
different motivation in \cite{khrip}.

Substituting the expression for $s_j$ one
easily gets the entropy to be \bb S=\ln N=\lambda A/2.\ee This means
that the Immirzi parameter has to be set at $\lambda /(2\pi)\simeq
0.274.$ Note that the summation over $s_j$ may raise this value
while the imposition of the zero projection constraint is expected
to lower it slightly.

The higher spins clearly raise the leading term, as in \cite{km},
but our expression is even larger than that of \cite{km}. The
difference arises from the fact that we have allowed all values $m=-j,...,j$
for all $j$, whereas \cite{km} did not distinguish states with the
same values of $m$ but different $j$. It is interesting to notice that their
equation 
\bb\sum_j2\exp [-\tilde\gamma\sqrt{j(j+1)}]=1,\ee which they got instead of
(\ref{lambda}), would have been obtained by us if we had restricted $m=\pm j$
for each $j$. This shows that although they wanted to count states
characterized by only the quantum numbers $m$ and satisfying
$2\sum\sqrt{|m|(|m|+1)}\leq A$, allowing for $|m|\leq j$,
their result is the same as though
they were interested only in states with $|m|=j$ and area {\em equal} to $A$.
States with lower values of $|m|$ appear to be negligibly fewer in comparison.

%We should also point out that if one wants to count states characterized by
%$m$, one ought to allow the value zero too. In that case, one has to take
%$m=\pm j$ for $j=1/2,3/2,2,5/2,...$ and $m=-1,0,+1$ for $j=1$. The correct
%equation for $\tilde\gamma$ is then
%\bb\sum_j2\exp [-\tilde\gamma\sqrt{j(j+1)}]+\exp [-\tilde\gamma\sqrt{2}]=1.\ee

Note further that if one allows $m$ to have all its $2j+1$ values for each
$j$, their first recursion relation (with the zero projection constraint
ignored) would get altered to \bb N(A)=\sum_j
(2j+1)N\Big(A-2\sqrt{j(j+1)}\Big)+\sqrt{A^2+1}\;,\ee which is
satisfied by our estimate $N(A)=\exp(\lambda A/2)$ with
$\lambda$ satisfying (\ref{lambda}) above. Our expression for the
entropy thus agrees with the solution obtained from the {\it
modified} recursion relation when the zero projection constraint is
ignored.

We shall now impose the constraint of zero angular momentum projection.
The number of configurations will be reduced somewhat, and a
correction is expected to emerge. Let $s_{j,m}$ punctures carry spin
$j$ and projection $m$, {\em i.e.} $s_j=\sum_ms_{j,m}$. Since at
each puncture $(j,m)$ assigns a unique state the total number of
states $N$ equals the number of ways $s_j$ and $s_{j,m}$ can be
distributed among themselves, \bb
N={(\sum_js_j)!\over\prod_js_j!}\prod_j{s_j!\over\prod_ms_{j,m}!}=
{(\sum_{j,m}s_{j,m})!\over\prod_{j,m}s_{j,m}!}\;,\label{njm}\ee
subject to the constraints $\sum_{j,m}ms_{j,m}=0$ and (\ref{areac}).
A lower bound is obtained by replacing $s_{j,m}$ for each $m$ by
$s_j/(2j+1)$ for the corresponding $j$. This
maximizes the number of combinations $s_j!/\prod_m s_{j,m}!$ for
each $j$ and also ensures zero total spin projection for each $j$,
hence for the sum. In Stirling's approximation, the main departure
from $(2j+1)^{s_j}$ occurs as the denominator contains a factor
$[s_j/(2j+1)]^{j+1/2}$, leading to a correction $-(j+1/2)\ln
[s_j/(2j+1)]$  (cf. \cite{lowerbound}) in $\ln N$. As $s_j/(2j+1)\propto
A\exp[-\lambda\sqrt{j(j+1)}]$, this correction can be expressed as 
\bb-\sum_j\Big[\ln A-\lambda\sqrt{j(j+1)}\Big] (j+1/2),\ee 
which appears to be
divergent. This happens because all $s_j$ have been assumed to be
large, although for large $j$, $s_j$ in the expression given above
goes to zero. So we restrict the sum to $j$ for which $s_j$ is
greater than unity. Taking the largest $j$ to be $n/2$, we see that
\bb\exp\Big[-\lambda \sqrt{n(n+2)/4}\Big] A\simeq 1,\ee so that \bb
n\simeq 2\ln A/\lambda.\ee Now \bb \sum_j j=n(n+1)/4\simeq (\ln
A)^2/\lambda^2.\ee Therefore the $\ln A$ piece  yields a $(\ln A)^3$
correction. The piece $-\lambda\sqrt{j(j+1)}$ also has to be taken
into account, using the sum \bb\sum_j j^2\simeq n^3/12\simeq 2 (\ln
A)^3/(3\lambda^3).\ee The total correction comes to $-(\ln
A)^3/(3\lambda^2)$: the total entropy is bounded by the contribution
of these configurations: \bb S\geq\lambda A/2 -(\ln
A)^3/(3\lambda^2).\ee This is our new lower bound.

It must be noted that this bound has been derived by assuming a
specific distribution of spins and spin projections to give the
largest number of combinations. Summing over different $s_j$ is
expected to increase the number of configurations. Note that there
also are additional nonleading terms in the expressions used above
which have been neglected, but these are much smaller in magnitude
than $(\ln A)^3$ and yield $(\ln A)^2$ and $\ln A$ pieces.

Let us now estimate the entropy, which as mentioned above is
expected to be higher than the above bound because of summation over
different configurations. In view of the zero spin projection
constraint, the number of configurations may be written by
explicitly summing over $s_{j,m}$ for each $j$ as (see
\cite{lowerbound}) \bb N_{\rm corr}={(\sum_j s_j)!\over \prod_j
s_j!}\int_{-2\pi}^{2\pi} {d\omega\over
4\pi}\prod_j\Big[\sum_{m_j}\exp (im_j\omega)\Big]^{s_j}.\ee This can
be rewritten as \bb N_{\rm corr}=\int_{-2\pi}^{2\pi}{d\omega\over
4\pi}N(\omega)\;,\ee where \bb N(\omega)= {(\sum_j s_j)!\over
\prod_j s_j!}\prod_j\Big [\sum_{m_j}\exp (im_j\omega)\Big]^{s_j}.\ee
To maximize $N_{\rm corr}$, we regard $s_j$ as functions
$s_j(\omega)$ subject to the area constraint and maximize
$N(\omega)$. The result is a simple modification of the one obtained
above, \bb N(\omega)=\exp(\lambda(\omega) A/2)\;,\ee where
$\lambda(\omega)$ satisfies \bb
1=\sum_j\exp\Big[-\lambda(\omega)\sqrt{j(j+1)}\Big]\sum_{m=-j}^j\exp\big(
i\omega m\big). \label{w}\ee This equation differs from that of
\cite{km} in $m$ going over $-j,...,j$, whereas their $m$ goes over
$\pm j$ as before. The modified recursion relation for $N(A,p)$,
which is the number of configurations satisfying the area constraint
(\ref{areac}) and the relation $\sum m=p$,
is \bb
N(A,p)=\sum_j\sum_{m=-j}^jN(A-2\sqrt{j(j+1)},p-m)+\theta\big(A-
2\sqrt{|p\/|(|p\/|+1)}\big),\ee
and gives rise to the above equation for $\lambda(\omega)$. 

For $\omega=0$, (\ref{w}) resembles (\ref{lambda}), so
$\lambda(0)=\lambda$. This yields the dominant contribution $\exp
(\lambda A/2)$ seen above. For small $\omega$, $\lambda(\omega)$
falls quadratically, and the $\omega$ integral becomes a gaussian,
which is readily seen to be proportional to $A^{-1/2}$  by
appropriate scaling. Thus,
\bb S_{\rm corr}=\ln N_{\rm corr}\sim\ln\big[{\exp(\lambda A/2)\over
A^{1/2}}\big]=\lambda A/2-\frac{1}{2}\ln A.\ee This is exactly
as in \cite{lowerbound,km}, indicating that the $(\ln A)^3,
(\ln A)^2$ terms do not survive when summed over configurations.

One can see this directly by approximating the sums over
configurations ({\em i.e.} sums over $s_{j,m}$) by integrals:
variation of $N$ in (\ref{njm}) with $s_{j,m}$ leads to a factor
$\exp [-(\delta s_{j,m})^2/2 s_{j,m}]$. Denominator factors of
$(2\pi s_{j,m})^{1/2}$ coming from Stirling's approximation are cancelled
by similar factors in the numerator coming from this gaussian
integration: \bb \int_{-\infty}^\infty d(\delta s_{j,m})\exp\Big
[-{(\delta s_{j,m})^2\over 2 s_{j,m}}\Big]=\big( 2\pi
s_{j,m}\big)^{1/2}.\ee Each  $s_{j,m}$ is proportional to $A$. The
area constraint and the spin projection constraint, which may be
thought of as reducing the number of summations, reduce the number
of factors of  $\sqrt{A}$ by two. But the numerator too has such a
factor through $(\sum s)^{1/2}$. An overall factor $1/\sqrt{A}$ is
thus left, as above, leading to the logarithmic correction with a
coefficient -1/2.

In conclusion, we have estimated the entropy of a black hole in the
quantum geometry approach by allowing spins of all non-zero values
at different punctures and regarding both $j$ and $m$ as relevant
quantum numbers. It was noted in \cite{ash} that the entropy
in the leading order is the same whether one considers $j$ as
relevant or not, with spin one-half assumed to yield the counting. However,
the dominant configuration, with the largest contribution 
to the number of states, contains spins higher than one-half, so that
the assumption made in \cite{ash} has to be relaxed.
The counting done in \cite{km} treated $j$ as irrelevant
and our result is different from theirs in the leading order,
although, somewhat surprisingly, the coefficient of the logarithmic 
term remains the same. The reason why $j$ has at times been 
disregarded in the state counting
is that this quantum number appears only in the volume Hilbert space
and {\em not} in the surface Hilbert space \cite{ash}, while it is the
surface Hilbert space which is considered to be the space of 
quantum states of the isolated horizon. However, although the area of a 
classical isolated horizon is defined intrinsically on the surface, the
area operator of a quantum isolated horizon is defined only to act on 
the volume Hilbert space. In fact,
the area of the horizon is determined by the `volume' quantum numbers $j$.
So, in our view, $j$ cannot be regarded as a hidden quantum number in
characterizing the states of a quantum isolated horizon. In the end, then,
the reason for the difference in our results
from \cite{km} is due to this difference in the definition of black hole
states. Which definition is more appropriate may be fixed either by making 
an independent estimate of the Immirzi parameter or by performing some
other semiclassical calculations from quantum isolated horizons.   
%The surprise is that the difference between the entropies in the
%two cases is so small. 

\end{document}